\documentclass[]{emulateapj}
\slugcomment{To appear in the Journal of The Korean Astronomical Society,
vol. 42, p. 17 (2009)}
\shorttitle{Mass Distribution in the Galactic Center}
\shortauthors{OH, KIM \& FIGER}

\def\spose#1{\hbox to 0pt{#1\hss}}
\newcommand\lsim{\mathrel{\spose{\lower 3.0pt\hbox{$\mathchar"218$}}
     \raise 2.0pt\hbox{$\mathchar"13C$}}}
\newcommand\gsim{\mathrel{\spose{\lower 3.0pt\hbox{$\mathchar"218$}}
     \raise 2.0pt\hbox{$\mathchar"13E$}}}
\newcommand\msun{{\rm \,M_\odot}}

\begin{document}
\title{Mass Distribution in the Central Few Parsecs of Our Galaxy}
\author{Seungkyung Oh$^{1,2}$, Sungsoo S. Kim$^{1,3,}$\altaffilmark{4}, and
Donald F. Figer$^{3}$}
\affil{$^{1 }$Dept. of Astronomy \& Space Science, Kyung Hee University,
Yongin-shi, Kyungki-do 446-701 \\
$^{2 }$Argelander-Institute for Astronomy, University of Bonn, Auf dem Huegel
71, 53121 Bonn, Germany \\
$^{3 }$Chester F. Carlson Center for Imaging Science, Rochester Institute
of Technology, 54 Lomb Memorial Drive, Rochester, NY 14623-5604, USA}
\altaffiltext{4}{Corresponding author}

\begin{abstract}
We estimate the enclosed mass profile in the central 10~pc of the Milky Way
by analyzing the infrared photometry and the velocity observations of
dynamically relaxed stellar population in the Galactic center.  HST/NICMOS
and Gemini Adaptive Optics images in the archive are used to obtain the number
density profile, and proper motion and radial velocity data were compiled
from the literature to find the velocity dispersion profile assuming a
spherical symmetry and velocity isotropy.  From these data, we calculate 
the enclosed mass and density profiles in the central 10~pc of the 
Galaxy using the Jeans equation.  Our improved estimates can better describe
the exact evolution of the molecular clouds and star clusters falling down
to the Galactic center, and constrain the star formation history of the inner
part of the Galaxy.
\end{abstract}


\section{INTRODUCTION}
\label{sec:intro}

The center of the Milky Way is the closest galactic nucleus, at a distance
of $\sim 8$~kpc from the Sun (Ghez et al. 2008; Gillessen et al. 2009),
and thus is a good laboratory to study galactic nuclei with.  Nonetheless,
the Galactic center (GC) had
not been studied well enough until 1990s, owing to large interstellar
extinction between the GC and the Sun ($A_{V} \gsim$ 30 mag; Rieke, Rieke,
\& Paul 1989), and the limit in the near-infrared (IR) observing technology
in the past.  Advances in high-resolution near-IR instruments during the
last two decades have yielded a wealth of information on the detailed
structure of the GC.

The most central part of the GC harbors a compact massive object, probably
a super massive black hole (SMBH) with a mass of $\sim 4 \times 10^6 \msun$
(Ghez et al. 2008; Gillessen et al. 2009).  Faint (compared to the stars
discussed right below) blue stars known as the ``S-stars'' or ``S-cluster''
are observed in the immediate vicinity (within 0.04~pc) of the SMBH.
Krabbe et al. (1995) identified them as massive main-sequence stars with a
spectral type of B0--B9.  Further from the center, between $\sim 0.04$
and $\sim 0.4$~pc, a few tens of OB supergiants, giants, and main-sequence
stars are observed along with a pool of faint red stars.  Paumard et al. (2006)
argue that these young stars form two disk-like orbital configurations,
highly inclined and rotating counter directions to each other, but note
also that Lu et al. (2009) find only one disk.

Kinematical properties of old GC population have been studied as well,
although not as intensively as for the young population.  Several studies
(e.g., Genzel et al. 1996, 2000; Figer et al. 2003) presented and analyzed
proper motion and radial velocity observations of the old stellar population.
This old population of stars is well relaxed under the influence of
SMBH and thus contains some information on the mass distribution
and dynamical environment around the SMBH. 

The presence of very young ($< 10^7$~yr) stars in the central parsec
has been a puzzle since the strong tidal forces and magnetic fields
in the GC as well as the elevated temperatures in molecular clouds
form hostile star formation environment there.  Inward migration
of a star cluster that is formed far outside the central parsec through
dynamical friction has been proposed (Gerhard 2001) to solve this youth
paradox, but it was shown that this scenario requires unrealistically
extreme initial cluster conditions to explain the observed distribution
of young stars in the central parsec (Kim \& Morris 2003; Kim, Figer, \&
Morris 2004).  Thus ``{\it in situ}'' star formation is a more likely
scenario, and star formation through a gravitationally unstable
gaseous disk around the SMBH (Nayakshin, Cuadra, \& Springel 2007, among
others) appears to be the most promising model currently.

The gas material that formed the young stars inside the central parsec
had probably come from the farther galactocentric distances, and a ring
of dense molecular gas extending 2--7~pc from the SMBH, called a
``circum-nuclear disk'' (CND; Christopher et al. 2005), is a good
candidate for its origin.  The CND itself shows some evidences of star
formation (Yusef-Zadeh et al. 2008) as well.

The initial configuration (e.g., radial and vertical sizes) of the
star-forming gaseous disk in the central parsec
will depend on the trajectory of the gas material infalling from the CND,
which in turn depends on the shape of the gravitational potential in that
region (i.e., the central few parsecs).  The exact shape of the potential
in the central few parsecs will also determine the degree of tidal
compression and distortion during the infall of the gas toward the central
parsec, which in turn will determine the star formation efficiency in the
central parsec.

The masses of the SMBH and its immediate vicinity ($< 0.1$~pc) have been
estimated by analyzing the line-of-sight (LOS) velocities and proper motions
of the S-stars (Ghez et al. 2008; Gillessen et al. 2009, among others), while
the enclosed mass profile (EMP) between 10 and 100~pc from the SMBH has
been calculated by analyzing the LOS velocities of the OH/IR stars (Lindqvist,
Habing, \& Winnberg 1992).  The EMP between $\sim 0.1$ and $\sim 5$~pc has
been studied either by interpreting the velocities of the CND as those of
a rotating, circular ring (Guesten et al. 1987; Jackson et al. 1993;
Christopher et al. 2005) or by using the LOS velocity dispersion of old,
relaxed stellar population (Genzel et al. 1996, 2000; Figer et al. 2003;
Sch\"odel et al. 2007).  The estimate from the former sensitively depends
on the assumption of a rotating, circular ring, and all of the latter
studies are based on the observations only out to $\sim 0.8$~pc
(Genzel et al. 2000 make use of the number count measurement out to $\sim 5$~pc
with the SHARP speckle camera on the 3.5-m New Technology Telescope [NTT]
obtained from the diploma thesis of Schmitt 1995, but the quality and
reliability of this measurement is difficult to be assessed).  Thus the EMP
estimate in the central few parsecs regime is still rather uncertain.

In the present paper, we estimate the EMP in this important region, between
$\sim 0.1$ and $\sim 10$~pc from the GC, by analyzing the {\it Hubble Space
Telescope} (HST) infrared photometry inside the central 5~pc along with the
radial velocities and proper motions of the old stellar population in the
same region.

\section{THE DATA}
\label{sec:data}

The Jeans equation for a spherically symmetric system will be used in the
present study to estimate the EMP in the GC, and for this, one needs number
density and velocity dispersion profiles of a relaxed population.
The former can be obtained from the stellar photometry, and the latter
from the proper motion observations and infrared spectroscopy.

\subsection{Stellar Photometry}

\begin{figure}
\epsscale{.5}
\plotone{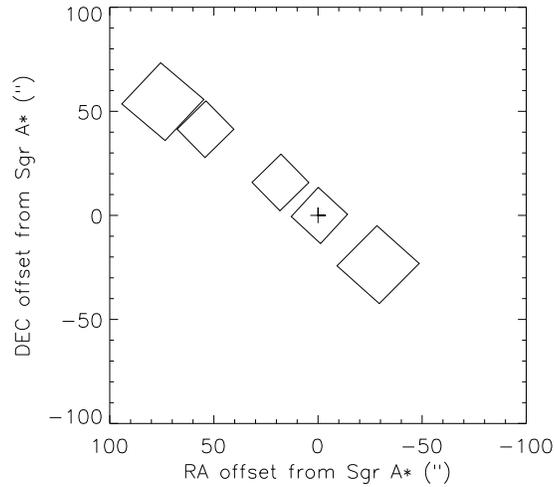}
\caption{\label{fig:frame}Locations and sizes of the {\it HST}/NICMOS images
used in the present study.  The cross at the center indicates the Sgr A$^*$.
The five frames are roughly aligned along the Galactic plane,
and the two larger frames are mosaiced ones each composed of 4 images.}
\end{figure}

Five near-IR images toward the GC taken with the NICMOS camera 2 (NIC2)
onboard the HST, which are available from the HST archive, have been
analyzed to obtain the stellar number density profile.  We adopted F160W and
F222M (similar to Johnson $H$ and $K$) filter images observed in October
1997 and September 2002.  Table \ref{table:frame} lists those image frames,
and Figure \ref{fig:frame} shows their sizes and locations, which cover
the central $\sim 4$~pc of the GC (100\arcsec =3.88~pc at the assumed
GC distance of 8~kpc).  The five frames are roughly aligned on the Galactic
plane, and the two larger frames are mosaiced ones each composed of 4 images.
The pixel scale and the field of view of each NIC2 image ($256 \times 256$
pixels) are $0.076 \arcsec$/pixel and $19.2 \arcsec \times 19.2 \arcsec$,
relatively.

We performed PSF photometry using the DAOPHOT package (Stetson 1987)
within the Image Reduction and Analysis Facility (IRAF).
7 to 10 bright and relatively isolated stars were used to construct a
point-spread function (PSF) of each image.  NICMOS PSFs have quite
prominent secondary diffraction rings and radial spikes, and automated
star-finding algorithms often identify the overlaps of two PSFs as stars.
We removed these bogus stars by hand.
We assume that the stars have the same intrinsic color $(m_{F160W}-
m_{F222M})_0= 0.25$~mag (see Kim et al. 2005) and calculate the reddening
of each star following an extinction law by Rieke et al. (1989).
The average extinction in our F222M image is found to be $\sim 3.2$ mag.

We carried out a completeness test by adding artificial stars to the
observed images.  We find that the 80~\% completeness limit of the
innermost F222M image is at 14 mag, and use stars brighter than 14 mag
for our analyses (the recovery fractions of the F222M images other than
the innermost one is greater than 90~\% at 14 mag).

Estimation of an EMP requires a density profile of a tracer population
that is dynamically relaxed.  Most of the stars outside the central 0.4~pc
of the GC are intermediate to old populations, while the stars inside
0.4~pc are a mixture of young and older populations (Krabbe et al. 1995).
To cull out the young population from the older, we use the CO line
strengths from the Adaptive Optics Demonstration Science Data Set of
the Gemini telescope.\footnote{Available at
http://gemini.conicyt.cl/{sciops/}{data/}{release\_doc/}{manual.html}}
The Gemini GC demo data were observed in July and August 2000 with $H$, $K'$,
$K$-continuum (centered at $2.26\,\micron$ with a bandpass of 60~\AA),
and CO (centered at $2.29\,\micron$ with a bandpass of 20~\AA) filters.
We performed PSF photometry for images 1 (roughly centered at the Sgr A$^*$)
and 2 (centered about 20\arcsec away from the Sgr A$^*$; each image has
a field of view of $20\arcsec \times 20\arcsec$).  We have calibrated
the photometry of the two images using the stars that appear on both
images, but they are not absolutely calibrated.

\begin{figure}
\epsscale{1.0}
\plotone{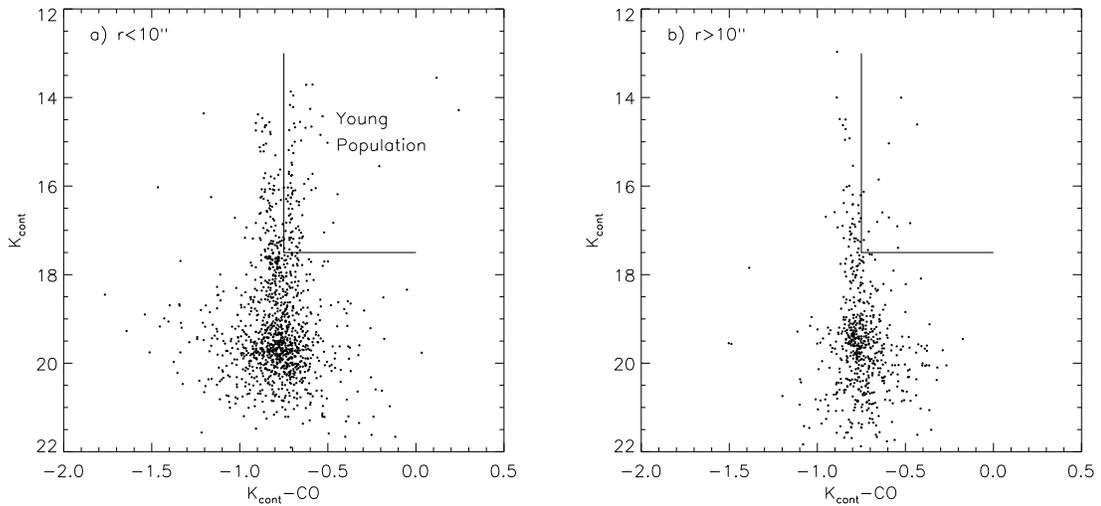}
\caption{\label{fig:gemini}$K_{cont}-$CO vs. $K_{cont}$ diagrams for
the stars in $r<10\arcsec$ (a) and $r>10\arcsec$ (b) regions from the
Gemini Adaptive Optics data.  The magnitudes presented here are not
calibrated ones (we find that our Gemini $K_{cont}$ magnitudes are
$\sim 3.6$~mag fainter than the NICMOS F222M magnitudes on average).
A separate population of stars is seen in the area defined by
$K_{cont}-{\rm CO}>-0.75$ and $K_{cont}<17.5$ (denoted with the straight
lines) at $r<10\arcsec$.  This population is mostly composed of young
stars (see the text).}
\end{figure}

Figure \ref{fig:gemini} shows $K_{cont}-$CO vs. $K_{cont}$ diagrams
for stars inside and outside the 0.4~pc radius from the Sgr A$^*$.  Outside
0.4~pc, the color-magnitude diagram forms a relatively narrow stream at
the bright end ($K_{cont}<16$), but inside 0.4~pc, a separate population
with redder $K_{cont}-$CO colors (i.e., smaller CO absorption strengths)
is seen.  The small CO strength is an indication of young ages
($\lsim 10^7$~yr), and we identify the stars in the box of Figure
\ref{fig:gemini}a as the young population.  We removed the young
stars in the central 0.4~pc from our NICMOS photometry by cross-identifying
the Gemini photometry against the NICMOS (the number of young stars
was 34 out of 1535 stars with ${\rm F222M}<14$ in our NICMOS photometry).
This way, we were able to have a collection of mostly intermediate to old
stars that are brighter than 14 mag in $K'$ in the central 4~pc of the GC,
and this sample will be used to produce the density profile of our tracer
population.

\subsection{Stellar Velocities}

Genzel et al. (2000) compiled a homogenized data set of stellar velocities
within the central 0.8~pc of the GC by combining various proper motion
and LOS velocity data from the New Technology Telescope (NTT), the Keck
telescope, and the MPG/ESO telescope on La Silla.  Figer et al. (2003)
reported LOS velocities of 85 cool stars in the central parsec of the GC
obtained with the Keck telescope.\footnote{Zhu et al. (2008) report the
second epoch observations of the LOS velocities toward the same region
with the same telescope, which mainly target the accelerations of the
stars.  The velocity data from the second epoch are very similar to the
first epoch, i.e., Figer et al. (2003).}  We combined these two data sets to
create a larger velocity sample of late-type (old) stars in the central parsec.
49 stars appear on both data sets, and we adopted the radial velocities
from the Keck for those common stars.  The total numbers of late-type
stars in this sample are 80 for the proper motion data and 236 for the LOS
velocity data.

Although not as much as found for young stars, old stars in the GC show
some figure rotation as well.  Following Genzel et al. (1996), we have
subtracted a rotational velocity of
\begin{equation}
\label{vrotz}
	v_{rot,z}=24(\Delta l / 5 \arcsec)^{0.4} \, {\rm km/s}
\end{equation}
from our LOS velocities ($\Delta l$ is the Galactic longitude
offset from Sgr A$^*$).

The stars in this data set extend out only to $\sim 0.8$~pc from
the Sgr A$^*$, we add to this data set two LOS velocity dispersion values
at $\sim 1.3$~pc and $\sim 4$~pc that Genzel et al. (1996) have obtained
from the literature (see their Table 6).

\section{NUMBER DENSITY PROFILES}
\label{sec:density}

\begin{figure}
\epsscale{1.0}
\plotone{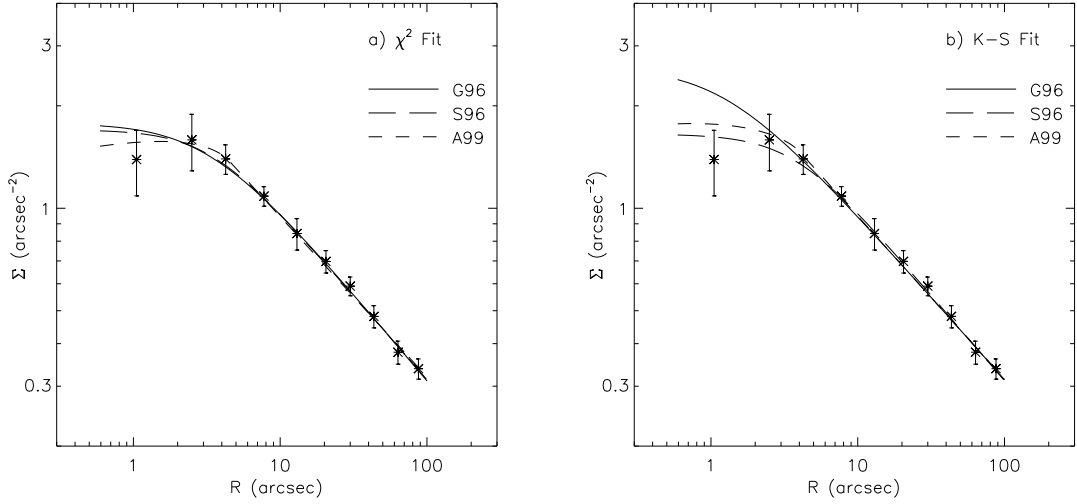}
\caption{\label{fig:sden}Best-fit surface density profiles for density
models of Genzel et al. (1996; G96), Saha et al (1996; S96), and Alexander
(1997; A99) using the $\chi^2$ test (a) and the Kolmogorov-Smirnov test (b).
The surface densities from our analysis of HST/NICMOS images (after the
subtraction of young population) are shown with asterisks and 1-$\sigma$
error bars.}
\end{figure}

Our photometry gives a distribution of projected stellar distances from
Sgr A$^*$ ($R$) for the tracer population in the central 4~pc region, and
the surface number density profile ($\Sigma[R]$) obtained from this
distribution is shown with asterisks in Figure \ref{fig:sden}.
To obtain a spatial number density profile ($n[r]$; $r$ is the spatial radius
from Sgr A$^*$) from $\Sigma(R)$, we model $n(r)$ with three functional forms
adopted in previous studies.

Genzel et al. (1996) implemented the following functional form for the spatial
density:
\begin{equation}
\label{ngenzel}
	n(r)=\frac{\Sigma_0}{r_0}\frac{1}{1+(\frac{r}{r_0})^\alpha}.
\end{equation}
Here, $\Sigma_{0}$, $r_{0}$, and $\alpha$ are the parameters to be found.
The information directly available from observations is $\Sigma(R)$,
and $n(r)$ can be transformed into $\Sigma(R)$ by the Abel integral equation,
\begin{equation}
\label{abel}
	\Sigma (R)=2\int_R^\infty\frac{n(r)}{\sqrt {r^2-R^2}}\, r \, dr.
\end{equation}
We numerically integrate this equation when fitting our $\Sigma(R)$ data
from the observations.  Genzel et al. (1996) fixed $\alpha$ to be an
asymptotic value at large $r$, 1.8, but here we keep it as a free parameter.
This density model will be designated as G96.

Saha, Bicknell, \& McGregor (1996) modeled the surface density with
\begin{equation}
\label{ssaha}
	\Sigma(R)=\Sigma_0\left(1+\frac{R^2}{R_0^2}\right)^{-\alpha},
\end{equation}
where $\Sigma_0$ and $R_0$ are to be fitted.  This model is a variant of
the modified Hubble law for elliptical galaxies, and has an analytical
counterpart for the spatial density of
\begin{equation}
\label{nsaha}
	n(r)=\frac{\Sigma_0}{R_0 B(1/2,\alpha)}\left(1+\frac{r^2}{R_0^2}
	                                       \right)^{-(\alpha+1/2)},
\end{equation}
where $B$ is the beta function. Saha et al. (1996) constrained $\alpha$ to
be 0.4, but we leave it as a free parameter.  This model will be called
S96.

Alexander (1999) adopted a broken power-law spatial density model:
\begin{equation}
\label{nalex}
	n(r)= \left\{\begin{array}{ll}
			n_0\left(r/r_0\right)^{-\alpha} \hspace{0.5cm} &
				\textrm{$r<r_0$} \\
			n_0\left(r/r_0\right)^{-\beta}  \hspace{0.5cm} &
				\textrm{$r\ge r_0 ,\  \beta >1$}
                     \end{array}\right.,
\end{equation}
where $n_0$, $r_0$, $\alpha$, and $\beta$ are the parameters to be fitted. 
Alexander (1999) assumed $r_0=0.4$~pc (10\arcsec) and $\beta=1.8$ for his
fits, but again, we keep them as free parameters.  We call this model A99.

When fitting the above functions to our NICMOS data, we try both
$\chi^{2}$ test and Kolmogorov-Smirnov (KS) test (see, e.g., Press et al.
1992 for the latter).  $\chi^2$ test is widely used as a standard statistical
method for comparing two distributions, but its result becomes sensitive
on the choice of the number and ranges of the bins if the number of
incidences in some of the bins is too small (say, less than $\sim 10$).
Since the number of intermediate to old stars in our data sample is small
at the very vicinity of the GC ($R<5"$), this may make the $\chi^2$ test
somewhat unreliable.  For this reason, we also use the KS test as a
supplementary test.  KS test utilizes the cumulative distribution function
instead of the histogram, and does not suffer the arbitrariness problem.
Note that, however, KS test has its own shortcoming, and this will be
discussed shortly.

Figure \ref{fig:sden} shows our best $\chi^2$ and KS fits of the above
density models to NICMOS data (Table \ref{table:sden} lists our best-fit
density model parameters).  The $\chi^2$ fits show a good agreement between
different density models although the innermost bin appears to be slightly
over-fitted.  The probabilities that the model has a different distribution
from the observation are less than 2~\% for all three models, so the
slight over-fit in the innermost bin is statistically not important.

The KS test results in similar fits for models S96 and A99, but it gives a
rather large discrepancy at the innermost bin for model G96.  This is probably
because (normalized) cumulative distribution functions always start
with 0 and end with 1, making the KS test rather insensitive at both ends
of the distribution.  Model G96 is the least flexible function
near the core radius ($r_0$), so it finds a bit difficult to adapt itself
to a sudden change near the core radius seen in the observation.
Nonetheless, all three KS fits are consistent with the observation by
better than 98~\% just as in $\chi^2$ tests.

We do not choose the best density model out of these six fits at this point.
Instead, we will see how much difference is made to the final EMP estimates
from these six models.

Note that all of our best-fits give $n \propto r^{-1.5}$ or similar
relations at large $r$.  This is somewhat shallower than those
by Sch\"odel et al. (2007; $r^{-1.75}$) and by Genzel et al. (2000;
$r^{-1.8}$).  These previous estimates are based on the photometry
with less sensitivity and/or smaller radii covered than in the present study,
and this is thought to be the cause of such differences.

\section{VELOCITY DISPERSION PROFILE}

\begin{figure}
\epsscale{1.0}
\plotone{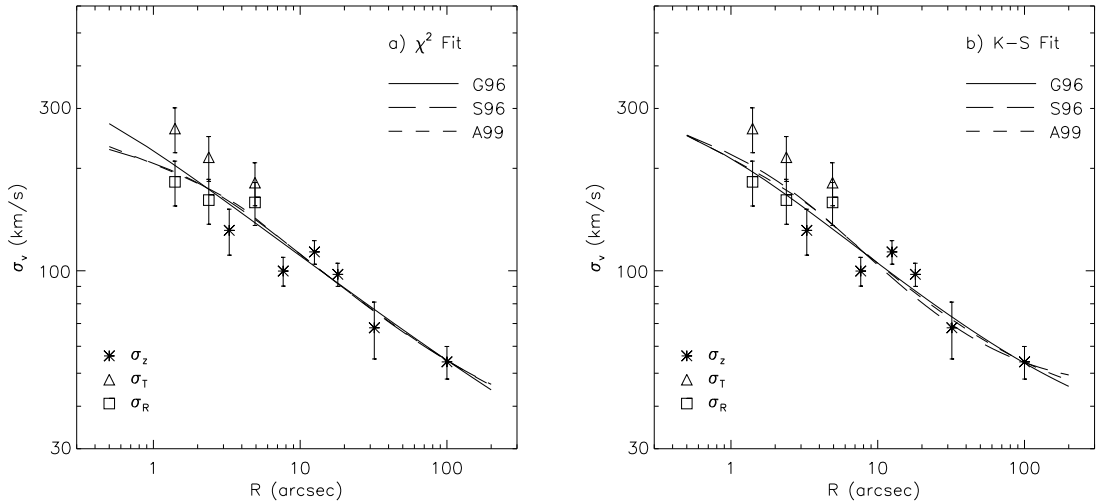}
\caption{\label{fig:velp}Best-fit velocity dispersion profiles for density
models of Genzel et al. (1996; G96), Saha et al (1996; S96), and Alexander
(1997; A99) using the $\chi^2$ test (a) and the Kolmogorov-Smirnov test (b).
The velocity dispersion data, which we obtained from Genzel et al. (1996, 2000)
and Figer et al. (2003), are shown with symbols and 1-$\sigma$ error bars.}
\end{figure}

The radii of the stars with proper motion data in our sample range from
$\sim 0.05$ to $\sim 0.5$~pc, and those with the LOS velocity data range
from $\sim 0.1$ to $\sim 5$~pc.  Thus our projected tangential and projected
radial velocity dispersions
($\sigma_T$ and $\sigma_R$, respectively) from the proper motion data
only cover the deepest region, and our LOS velocity dispersion ($\sigma_z$)
covers much wider region except the innermost area.  These dispersion
profiles are shown with three different symbols in Figure \ref{fig:velp}.

For a functional form of spatial (i.e., not projected) velocity dispersion
($\sigma_v$), we adopt the following parameterization used by Genzel et al.
(1996):
\begin{equation}
\label{sigmar}
	\sigma_v(r)^2 = \sigma_\infty^2+\sigma_0^2(r/r_{0v})^{-\gamma}.
\end{equation}
For $r_{0v}$, we use the same $r_0$ or $R_0$ of the density model that
is used to fit the above equation to the observed velocity dispersions
(see below).  Thus only $\sigma_\infty$, $\sigma_0$, and $\gamma$ are
the parameters to be found.

We assume that the velocity distribution is isotropic.  We do not try
anisotropic velocity models in the present paper because the spatial
coverages of our proper motion sample and LOS velocity sample overlap
only marginally.

Since we assume an isotropy for the velocity dispersion, the relation
between the observed velocity dispersion and the spatial velocity
dispersion is given by
\begin{eqnarray}
\label{sigmap}
	\sigma_T(R)^2 & = \sigma_R(R)^2 = \sigma_z(R)^2
	              & = \frac{2}{\Sigma(R)} \int_R^\infty
                          \frac{n(r)\sigma_v(r)^2}{\sqrt{r^2-R^2}}\, r \, dr.
\end{eqnarray}
Thus obtaining a functional form for $\sigma_v(r)$ profile requires
spatial and surface density profile information, and we try all of our
six fits for $\Sigma(R)$ (and its corresponding $n[r]$) obtained in the
previous section.  Figure \ref{fig:velp} plots our model (eq. \ref{sigmap})
fits to the observed velocity dispersions with three different density
profiles (G96, S96, and A99) and two different statistical tests ($\chi^2$
and KS).  All six fits result in very similar velocity dispersion profiles
(best-fit velocity dispersion parameters are listed in Table \ref{table:vdisp}).

\section{ENCLOSED MASS PROFILE IN THE GALACTIC CENTER}

The Jeans equation is the first moment of the collisionless Boltzmann
equation, and it gives a relation between the enclosed mass of a system
and the velocity dispersions.  For a spherically symmetric, rotating system
with an isotropic velocity distribution, the equation becomes
\begin{equation}
\label{jeans}
	\frac{GM(r)}{r} = v_{rot}(r)^2+\sigma_v(r)^2
			  \left\{ -\frac{d\ln[n(r)]}{d\ln r}
				  -\frac{d\ln[\sigma_v(r)^2]}{d\ln r}
			  \right\},
\end{equation}
where $M$ is the enclosed mass and $v_{rot}$ is the rotational velocity.
The latter as a function of $r$ can be obtained from the observed, mean
LOS velocities as a function of the longitude offset from Sgr A$^*$ by
the following relation: $v_{rot,z}(\Delta l)$:
\begin{equation}
\label{vrot}
	v_{rot}(r) = -\frac{r}{\pi n(r)} \int_r^\infty
		      \frac{d}{d\Delta l} \left (
			  \frac{\Sigma(\Delta l) v_{rot,z}(\Delta l)}{\Delta l}
		      \right ) \frac{d \Delta l}{\sqrt{\Delta l^2-r^2}}.
\end{equation}
For the density profile in this Abel transform, we use the density model
for old population in Genzel et al. (1996) instead of those obtained in the
present study because it better represents the whole inner bulge (several tens
to hundreds of parsecs) in which the figure rotation takes place.

\begin{figure}
\epsscale{1.0}
\plotone{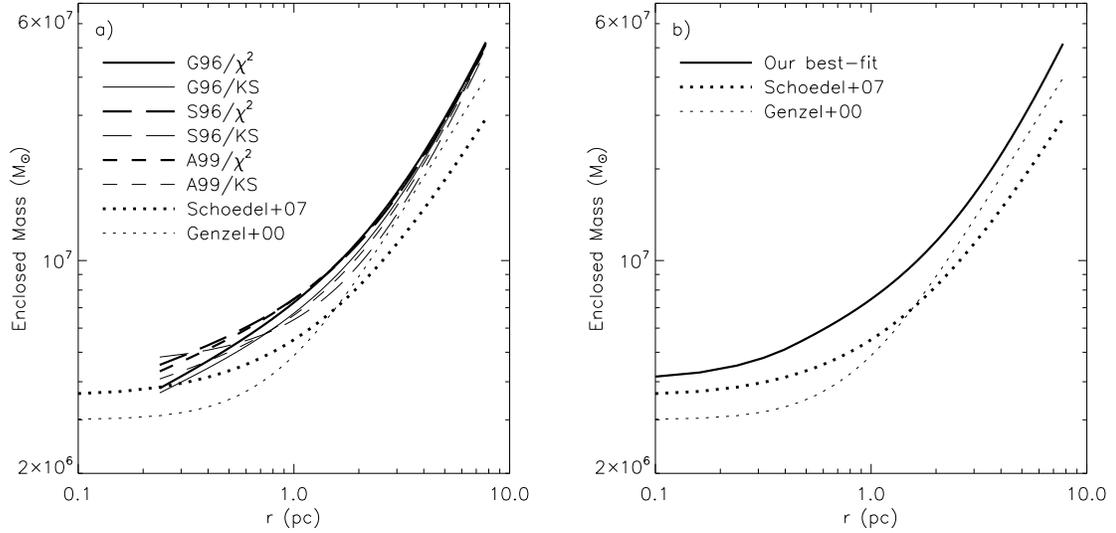}
\caption{\label{fig:mass}(a) Enclosed mass profiles from the Jeans equation
for density models of Genzel et al. (1996; G96), Saha et al (1996; S96),
and Alexander (1999; A99) and statistical tests of $\chi^2$ and
Kolmogorov-Smirnov (KS) methods.
(b) Enclosed mass profile of A99/$\chi^2$ model modified to converge
to the SMBH mass of $4 \times 10^6 \msun$ at $r=0$~pc (our best-fit).
Also plotted are the enclosed mass profiles estimated by Sch\"odel et al.
(2007) and Genzel et al.
(2000).}
\end{figure}

Figure \ref{fig:mass}a shows the EMPs obtained from the above Jeans
equation with our density and velocity dispersion fits for the intermediate
to old stellar populations.  These profiles are quite similar to each other,
but quite different from the two previous studies, Genzel et al. (2000) and
Sch\"odel et al. (2007).  Our EMP is similar to the former at larger radii
but is considerably larger than the former.  The latter has much
smaller EMP in the larger radii because it is based on a rather simple
approximation that the LOS velocity dispersion in the outer region is
constant.

We do not plot our EMPs inside 0.2~pc.  Our EMP estimates at this inner region
are not reliable because there are not many bright, intermediate to old stars
in this region (this ``hole'' of old stars is probably due to frequent close
encounters with other stars; see Genzel et al. 1996).

For a smooth convergence of the EMP to the mass of the SMBH, we extrapolate
the EMP inward from the radius at $M=5 \times 10^6 \msun$ assuming that the
extended mass (the mass excluding the SMBH) distribution follows a power-law
function and that the mass of the SMBH is $4 \times 10^6 \msun$ (Ghez et al.
2008; Gillessen et al. 2009).  Figure \ref{fig:mass}b (and Table
\ref{table:profile}) shows such inward extrapolation of our $\chi^2$ best-fit
of A99 density model (A99/$\chi^2$).  We choose this model as our canonical fit
as it describes the observed density profile the best and its EMP roughly
represnts the average of all our 6 models.

\begin{figure}
\epsscale{0.5}
\plotone{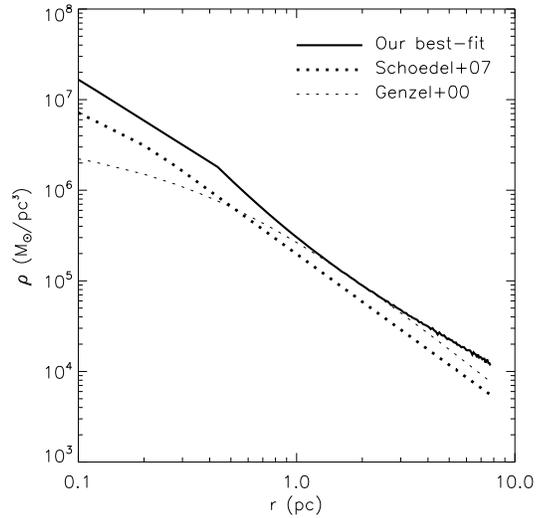}
\caption{\label{fig:rho}Density profile of the modified A99/$\chi^2$ model
(our best-fit; see Fig. \ref{fig:mass}b) along with those estimated by
Sch\"odel et al. (2007) and Genzel et al. (2000).}
\end{figure}

Local densities of the extended mass (the mass other than the SMBH) can be
obtained by differentiating the EMP.  The density profile for our A99/$\chi^2$
model is given in Figure \ref{fig:rho} and Table \ref{table:profile}.
Again, our result is quite close to that of Genzel et al. at larger radii,
but closer to that of Sch\"odel et al. at smalle radii.

\section{SUMMARY AND DISCUSSION}

We have estimated the EMP in the central 10~pc of the Milky Way
by analyzing the infrared photometry and the velocity observations of
dynamically relaxed stellar population in the Galactic center.  
HST/NICMOS and Gemini images were used to obtain the number
density profile of the relaxed population, and the LOS velocities and
proper motion data of the same population were used to calculate the EMP
from the Jeans equation assuming a spherical symmetry and velocity isotropy.
The newly obtained EMP is larger than the previous studies at
$0.1 < r < 10$~pc, which is consistent with the most recent value
for the mass of the SMBH.

As discussed in \S 1, the enclosed mass and density profiles in the central
few parsecs will determine the exact morphological evolution of the molecular
clouds and star clusters that are moving down to the GC.  Our larger EMP
implies that star clusters and
molecular clouds in the central few parsecs will have smaller tidal radii
and shorter orbital periods than previously expected.

Degenerate stellar remnants such as stellar mass black holes and neutron
stars are thought to be segregated in the GC due to their gradual dynamical
friction with lighter field stars (Morris 1993).  Our density profile
can be used to constrain the amount of segregated degenerate objects.
Since the current reservoir of degenerate stars is a result of continuous
star formation in the GC (Figer et al. 2004), a detailed study on the
stellar populations in the GC based on our EMP estimate will constrain
the star formation history in the GC as well.

The anonymous referee suggested density profile functions that have
an inner radial cutoff to better describe the ``hole'' of old stars.
We do not try such functions here because they will not significantly
change our results on the EMP and the nature of the hole is beyond
the scope of the present paper, but we do believe that such functions
can give some constraints on the size of the hole.

\acknowledgements
This work was supported by the 2009 Sabbatical Leave program of Kyung Hee
University.  S. O. was supported by the Astrophysical Research Center for the
Structure and Evolution of the Cosmos (ARCSEC) of the Korea Science and
Engineering Foundation through the Science Research Center (SRC) program.
The material in this paper is partially supported by NASA under award NNG
05-GC37G, through the Long Term Space Astrophysics program. The research
by D. F. F. was performed in the Rochester Imaging Detector Laboratory with
support from a NYSTAR Faculty Development Program grant.)


\clearpage
\begin{deluxetable}{crc}
\tablecolumns{3}
\tablewidth{0pt}
\tablecaption{
\label{table:frame}HST/NICMOS Datasets Analyzed in the Present Study}
\tablehead{
\colhead{Dataset Name} &
\colhead{Exposure (Sec)} &
\colhead{Observation Date}
}
\startdata
\sidehead{F160W}
N49Z010H0  &   25.9  &  October 1997    \\
N49Z01030  &   55.9  &  October 1997    \\
N49Z01070  &   55.9  &  October 1997    \\
N49Z010B0  &   55.9  &  October 1997    \\
N49Z010F0  &   55.9  &  October 1997    \\
N49Z02030  &   55.9  &  October 1997    \\
N49Z02070  &   55.9  &  October 1997    \\
N49Z020B0  &   55.9  &  October 1997    \\
N49Z020F0  &   55.9  &  October 1997    \\
N6LO02050  &   24.0  &  September 2002  \\
N6LO03050  &   24.0  &  September 2002  \\
\sidehead{F222M}
N49Z010I0  &   71.9  &  October 1997    \\
N49Z01040  &  207.9  &  October 1997    \\
N49Z01080  &  207.9  &  October 1997    \\
N49Z010C0  &  207.9  &  October 1997    \\
N49Z010G0  &  207.9  &  October 1997    \\
N49Z02040  &  207.9  &  October 1997    \\
N49Z02080  &  207.9  &  October 1997    \\
N49Z020C0  &  207.9  &  October 1997    \\
N49Z020G0  &  207.9  &  October 1997    \\
N6LO02040  &   71.9  &  September 2002  \\
N6LO03040  &   71.9  &  September 2002  \\
\enddata
\end{deluxetable}

\begin{deluxetable}{clcclcclcclcclccl}
\tablecolumns{17}
\tablewidth{0pt}
\tablecaption{
\label{table:sden}Surface Number Density Fits}
\tablehead{
\multicolumn{2}{c}{G96} &
\colhead{} &
\multicolumn{2}{c}{S96} &
\colhead{} &
\multicolumn{2}{c}{A99} &
\colhead{} &
\multicolumn{2}{c}{G96} &
\colhead{} &
\multicolumn{2}{c}{S96} &
\colhead{} &
\multicolumn{2}{c}{A99}
}
\startdata
\multicolumn{8}{c}{$\chi^2$ Fit} & \hspace{0.5cm} & \multicolumn{8}{c}{KS Fit} \\ \cline{1-8} \cline{10-17}
$\Sigma_0$ &  247 & & $\Sigma_0$ & 1130 & & $n_0$    & 994.7 & & $\Sigma_0$ &  343 & & $\Sigma_0$ & 1100 & & $n_0$    & 1168.1 \\
$r_0$      & 0.11 & & $R_0$      & 0.13 & & $r_0$    & 0.17 & & $r_0$      & 0.44 & & $R_0$      & 0.14 & & $r_0$    & 0.15 \\
$\alpha$   & 1.50 & & $\alpha$   & 0.25 & & $\alpha$ & 0.45 & & $\alpha$   & 1.48 & & $\alpha$   & 0.25 & & $\alpha$ & 0.59 \\
           &      & &            &      & & $\beta$  & 1.47 & &            &      & &            &      & & $\beta$  & 1.48 \\
\enddata
\tablecomments{$\Sigma_0$ are in units of pc$^{-2}$, $n_0$ in pc$^{-3}$, and $r_0$ \& $R_0$ in pc.}
\end{deluxetable}

\begin{deluxetable}{ccccccccc}
\tablecolumns{9}
\tablewidth{0pt}
\tablecaption{
\label{table:vdisp}Velocity Dispersion Profile Fits}
\tablehead{
\colhead{} &
\colhead{} &
\colhead{G96} &
\colhead{S96} &
\colhead{A99} &
\colhead{} &
\colhead{G96} &
\colhead{S96} &
\colhead{A99}
}
\startdata
 & \hspace{0.5cm} & \multicolumn{3}{c}{$\chi^2$ Fit} & \hspace{0.5cm} & \multicolumn{3}{c}{KS Fit} \\ \cline{3-5} \cline{7-9}
$\sigma_\infty$ & & 30.9 & 31.3 & 28.1 & & 44.9 & 44.8 & 38.0 \\
$\sigma_0$      & & 253  & 237  & 209  & & 170 & 201  & 210  \\
$\gamma$        & & 0.77 & 0.77 & 0.73 & & 1.11 & 0.99 & 0.81 \\
\enddata
\tablecomments{$\sigma_\infty$ and $\sigma_0$ values are in units of km/s.}
\end{deluxetable}

\begin{deluxetable}{rcc}
\tablecolumns{3}
\tablewidth{0pt}
\tablecaption{
\label{table:profile}Enclosed Mass and Density Profiles in the Galactic Center}
\tablehead{
\colhead{$r$} &
\colhead{$M$} &
\colhead{$\rho$}\\
\colhead{(pc)} &
\colhead{($\msun$)} &
\colhead{($\msun$/pc$^3$)}
}
\startdata
 0.1 & 4.15e+06 & 2.49e+07 \\
 0.2 & 4.40e+06 & 5.87e+06 \\
 0.3 & 4.73e+06 & 3.25e+06 \\
 0.5 & 5.54e+06 & 1.33e+06 \\
 0.7 & 6.33e+06 & 6.30e+05 \\
 1.0 & 7.47e+06 & 3.04e+05 \\
 2.0 & 1.16e+07 & 8.88e+04 \\
 3.0 & 1.65e+07 & 4.76e+04 \\
 5.0 & 2.90e+07 & 2.29e+04 \\
 7.0 & 4.48e+07 & 1.42e+04 \\
\enddata
\end{deluxetable}

\clearpage

\end{document}